\documentclass{article}
\usepackage[top=0.85in,left=1.25in,footskip=0.75in]{geometry}

\usepackage{changepage}

\usepackage[utf8]{inputenc}

\usepackage{textcomp,marvosym}

\usepackage{fixltx2e}

\usepackage{amsmath,amssymb}
\usepackage{graphicx}

\usepackage{cite}

\usepackage{nameref,hyperref}

\usepackage[right]{lineno}

\makeatletter
\renewcommand{\@biblabel}[1]{\quad#1.}
\makeatother

\date{}

\usepackage{lastpage,fancyhdr,graphicx}
\usepackage{epstopdf}
\DeclareMathOperator{\atan}{atan}

\DeclareMathOperator{\atanh}{atanh}

\begin{document}

%\markboth{John H Marr}{Visualizing the photon path in SR}

\title{A Novel Visualization of the Geometry of Special Relativity}
\maketitle
\begin{center}
John H Marr\\
{\it Unit of Computational Science \\
Building 250, Babraham Research Campus \\
Cambridge, CB22 3AT, UK.}\\
john.marr@2from.com
\end{center}

Submitted to Int. J. Mod. Phys. C (3 Sept 2015)

\begin{abstract}
The mathematical treatment and graphical representation of Special Relativity (SR) are well established, yet carry deep implications that remain hard to visualize. This paper presents a new graphical interpretation of the geometry of SR that may, by complementing the standard works, aid the understanding of SR and its fundamental principles in a more intuitive way. From the axiom that the velocity of light remains constant to any inertial observer, the geodesic is presented as a line of constant angle on the complex plane across a set of diverging reference frames. The resultant curve is a logarithmic spiral, and this view of the geodesic is extended to illustrate the relativistic Doppler effect, time dilation, length contraction, the twin paradox, and relativistic radar distance in an original way, whilst retaining the essential mathematical relationships of SR. Using a computer-generated graphical representation of photon trajectories allows a visual comparison between the relativistic relationships and their classical counterparts, to visualize the consequences of SR as velocities become relativistic. The model may readily be extended to other situations, and may be found useful in presenting a fresh understanding of SR through geometric visualization.  

\end{abstract}
%%%%%%%%%%%%%%%%%%%%%%%%%%%%%%%%%%%%%%%%%%%%
\section{Introduction}
Visualization is an important tool in developing models of science, and particularly relativity,\cite{Ruder2008} because although the mathematical treatment of special relativity (SR) is well established, its strong geometric component lends itself to visualization without relying on symbolic notation.\cite{Weiskopf2010}  
Graphical representation of Einstein's defining work on SR was first provided by 
Minkowski,\cite{Minkowski1909} and current teaching on SR has remained essentially unchanged since Weyl's 
original book showing the light-cone.\cite{Weyl1922} 
Much subsequent teaching has been based on the figures in his book, and variations of these 
diagrams usually form the definitive background for presenting SR (e.g. Ref.~\cite{Sartori1996}). 
However, Minkowski diagrams and their associated hyperbolas do not present a direct 
visualization of Einstein's premise that the vacuum speed of light is a universal constant for 
any observer in an inertial reference frame, regardless of their own motion or that of the 
light source.

This paper presents an original approach to the geometry of SR using the core axiom that the velocity of light in a vacuum, $c$, is identical to every inertial observer. The photon paths are then constructed, consistent with the standard view of the photon as lying along the surface of a light cone centred on any given event along an arbitrary world-line. 
In particular, any two connected events, such as the emission and detection of a photon, 
may be represented by two inertial world-lines that intersect at a common origin, and between these world-lines a set of intervening world-lines may be constructed through the origin, on each of which the photon path may be drawn as a purely local light cone.
By equating this set of world-lines with rotation on the complex plain, it is demonstrated that the photon path between two such connected events draws a line of constant angle to this set of radii.

From this concept, a geometrical model is constructed to show the geodesic of the photon as the locus of an equiangular line, or logarithmic spiral, to this set of inertial frames. 
This alternative view of the geodesic as a logarithmic spiral can be used to demonstrate some common concepts of SR such as the relativistic Doppler effect, time dilation, the twin paradox, and relativistic radar distance, in a more intuitive way to complement the standard textbooks, and emphasis is given to observable differences in measurement between these relativistic relationships and their classical counterparts.
%%%%%%%%%%%%%%%%%%%%%%%%%%%%%%%%%%%%%%%%%%%%
\section{The geometry of Special Relativity}
%%--------------------
\begin{figure}[ht]
   \centering
   \includegraphics[width=4cm]{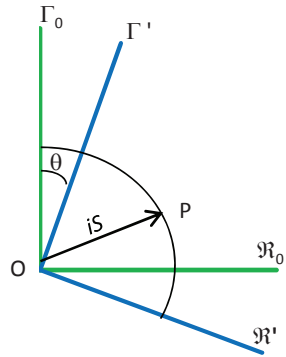}
   \caption{Invariance of the interval $iS$ under rotation.}
   \label{fig:rotation}
	\vspace*{8pt}
\end{figure}
%%--------------------
The Universe may be considered as essentially geometrical in its description of the position and motion of objects in space and time. 
In terms of the three space dimensions, $x_1$, $x_2$, $x_3$, we may define the sphere of radius $R$ 
and, as readily shown in standard works (e.g.~\cite{Poincare1904, Muirhead1973}), 
by considering a photon travelling at $c$ it is possible to treat the time axis as a fourth coordinate, $x_0=ict=\tau$, 
where $i=\sqrt{-1}$. This leads to the definition of an interval as the equation of a four-dimensional sphere 
(the spherime) of radius $iS$ (Eq.~\ref{eq:1}).
\begin{equation}
-S^2=x_0^2+x_1^2+x_2^2+x_3^2=\sum_{i=0}^3x_i^2\,.
\label{eq:1}
\end{equation}
This leads naturally to the spatialization of time as proposed by Minkowski, 
and in common with the radius of the circle and sphere, S remains invariant under rotation.\cite{Poincare1904,Minkowski1907}
This is illustrated in Fig.~\ref{fig:rotation} where the reference frame $\Sigma_0$, with real axis $\Re_0$ and imaginary axis $\Gamma_0$, is rotated through $i\theta$ to the reference frame $\Sigma'$ 
with axes $\Re'$ and  $\Gamma'$, with $\Re_0$ and $\Re'$ taken to be the three spatial axes compacted to one dimension.
This is justified because any two lines sharing a common origin may be considered to lie on a plain.
$\Sigma'$ now represents a reference frame moving with constant velocity $v$ with respect to $\Sigma_0$. 
The radius $OP$ (corresponding to the interval $iS$) remains invariant, and it may readily be shown that the following transformation rules apply:
\begin{equation}
R'=R_0\cos{(i\theta)}-\Gamma_0\sin{(i\theta)}\,.
\label{eq:2}
\end{equation}
\begin{equation}
\Gamma'=R_0\sin{(i\theta)}+\Gamma_0\cos{(i\theta)}\,.
\label{eq:3}
\end{equation}
\begin{equation}
R_0=R'\cos{(i\theta)}+\Gamma'\sin{(i\theta)}\,.
\label{eq:4}
\end{equation}
\begin{equation}
\Gamma_0=-R'\sin{(i\theta)}+\Gamma'\cos{(i\theta)}\,.
\label{eq:5}
\end{equation}
If now we consider rotation of the axes so that $OP$ lies on the $\Gamma_0$ axis, 
then $R_0=0$, and $R'\cos{(i\theta)}+\Gamma'\sin{(i\theta)}=0$.~Hence
\begin{equation}
\tan{(i\theta)} = -R'/\Gamma' = -R'/{ict'}=iv/c = i\beta\,.
\label{eq:6}
\end{equation}
where $\beta=v/c$. And by Pythagoras:
\begin{equation}
\cos{(i\theta)}=c/{(c^2-v^2)^{1/2}}=1/{(1-v^2/c^2)^{1/2}}=\gamma\,.
\label{eq:7}
\end{equation}
\begin{equation}
\sin{(i\theta)}=-iv/{(c^2-v^2)^{1/2}}=i\beta\gamma\,.
\label{eq:8}
\end{equation}
Where $\gamma$ is the Lorentz factor. Substitution of (\ref{eq:7}) and (\ref{eq:8}) back into (\ref{eq:4}) and (\ref{eq:5}) may readily be seen to yield the Lorentz transformation equations:
\begin{equation}
R_0=\gamma(R'-vt')\,.
\label{eq:9}
\end{equation}
\begin{equation}
t_0=\gamma(t'-vR'/c^2)\,.
\label{eq:10}
\end{equation}
Alternatively, as $\tan{(i\theta)}=i\tanh{\theta}$, $\sin{(i\theta)}=i\sinh{\theta}$ 
and $\cos{(i\theta)}=\cosh{\theta}$,  we may write:
\begin{equation}
\beta=\tanh{\theta}\,.
\label{eq:11}
\end{equation}
\begin{equation}
\beta\gamma=\sinh{\theta}\,.
\label{eq:12}
\end{equation}
\begin{equation}
\gamma=\cosh{\theta}\,.
\label{eq:13}
\end{equation}

%%%%%%%%%%%%%%%%%%%%%%%%%%%%%%%%%%%%%%%%%%%%
\subsection{Constructing the photon path}
%%--------------------
\begin{figure}[ht]
   \centering
   \includegraphics[width=7cm]{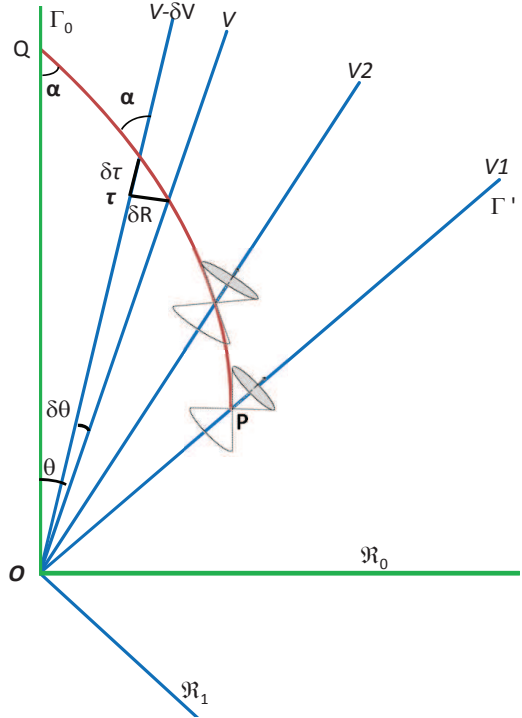}
   \caption{Geodesic for photon traversing frames moving at different velocities 
   relative to $\Sigma_0$, and a small element of the geodesic $\delta{}S$ for diverging reference frames. 
   Also shown are two light cones in their local frames of reference.}
   \label{fig:geodesic}
\end{figure}
%%--------------------
Let $\Sigma_0$ be the inertial frame of an observer with time and space axes $\Gamma_0$, $\Re_0$, and let $\Sigma'$  be a diverging 
inertial frame with axes $\Gamma'$, $\Re'$, moving with uniform velocity $V1$ relative to $\Sigma_0$, that was coincident with 
$\Sigma_0$ at $O$ (Fig.~\ref{fig:geodesic}). In physical terms, a point moving along $\Gamma'$ is moving at constant velocity w.r.t. the 
$\Sigma_0$ frame, but is stationary within its own $\Sigma'$ frame. Suppose such a point emits a photon ($P$, Fig.~\ref{fig:geodesic}). 
This will be at time $\tau'$ along its time axis $\Gamma'$, and it will be observed in $\Gamma_0$ at time $\tau_0$, measured at Q along 
$\Gamma_0$. The photon will leave $\Gamma'$ at velocity $c$ tangential to the local light cone at $P$, but because $c$ is constant to 
all observers, and all reference frames through $O$ are equivalent under rotation, this line will trace a geodesic through spacetime 
such that it is tangential to the light cone of every rotated reference frame such as $V2$. The locus of this line is shown by the red 
line, at a constant angle $\alpha$ to all lines passing through $O$. By convention the scale is chosen such that $c=1$ with 
$\alpha=45^{\circ}$.  It must be emphasised that the intervening world-lines between $\Gamma'$ and $\Gamma_0$ are purely hypothetical, 
and the geodesic is a mathematical abstraction. In reality, we know only the two events $P$ and $Q$, and although the geodesic 
represents the shortest path between them in analogy to a great circle, in practice we know nothing of the actual path of the photon. 
It should be noted that light cones represent proper (local, or ``real'') time with trigonometric angles, whereas because the 
rotated time axes of Fig.~\ref{fig:geodesic} are the imaginary axes $\tau=ict$, the angle of rotation $\theta$ corresponds to the 
hyperbolic functions rather than trigonometric ones and may be greater than $45^{\circ}$. Also, whereas the rotation of 
Fig.~\ref{fig:rotation} demonstrates the transformation rules with constant radii, the radii of Fig.~\ref{fig:geodesic} represent 
increasing time on the photon's path between $P$ and $Q$.

Fig.~\ref{fig:geodesic} also shows a small element of the photon's path moving a distance $\delta{}R$ 
over a time element $\delta\tau$, between two frames diverging from each other by a small velocity $\delta{}v$. 
Such an element is considered to be at a time $\tau$ from $O$, and subtends an angle $\delta\theta$ with $O$. 
For a photon, $\delta{}S=0$. It then follows:
\begin{equation}
\delta{}R^2+\delta\tau^2=0 \textnormal{ (null geodesic for photon)}
\label{eq:14}
\end{equation}
\begin{equation}
\textnormal{Hence }\delta{}R^2=-\delta\tau^2=c^2\delta{}t^2 
\textnormal{~or~}\delta{}R=\pm c\delta{}t\,,
\label{eq:15}
\end{equation}
where the sign represents an incoming or outgoing photon. But $\delta{}R=\tau\delta{}i\theta$ and $\tau=ict$, hence:
\begin{equation}
c\delta{}t=ict\delta{}i\theta \textnormal{, or~} \delta{}t/t=\mp \delta\theta\,.
\label{eq:16}
\end{equation}
Using $-\delta\theta$ for the incoming photon and integrating:
\begin{equation}
\int_{t'}^{t_0}\frac{\textnormal{d}{t}}{t}=\int_\theta^0-\textnormal{d}\theta\,.
\label{eq:17}
\end{equation}
\begin{equation}
\textnormal{i.e. } \ln{(t_0/{t'})}=\theta \textnormal{ or } t_0/{t'}=e^{\theta}\,.
\label{eq:18}
\end{equation}
which is the equation of the logarithmic spiral on the imaginary plane. 
It may be noted that the geometry allows $\theta>360^{\circ}$ 
because $v/c =\tanh{\theta}\rightarrow 1$ 
as $v\rightarrow c$ and $\theta\rightarrow \infty$.

%%%%%%%%%%%%%%%%%%%%%%%%%%%%%%%%%%%%%
\section{Relativistic Doppler effect}
In physical terms, all diverging world-lines are equivalent and will ``see'' the photon intercepting them 
at velocity $c$, and leaving them at velocity $c$. 
When the photon arrives at $\Gamma_0$, its actual source line is indistinguishable in terms of $v$ 
(this is Einstein's second postulate); however, the source lines are not completely indistinguishable, 
because diverging reference frames are Doppler red-shifted.

\begin{figure}[ht]
   \centering
   \includegraphics[width=5cm]{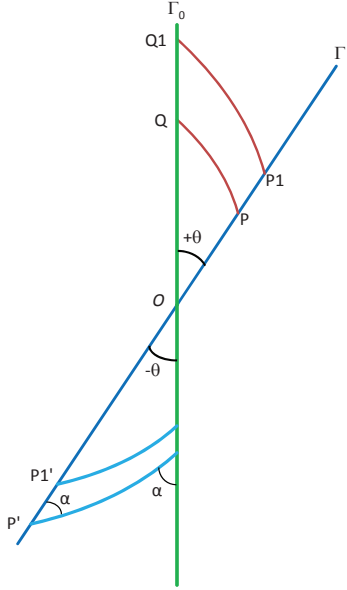}
   \caption{Relativistic Doppler effect. P-P1 lie on $\Gamma'$ receding from $O$ with uniform velocity, 
   with redshifted photons received on $\Gamma_0$ at Q-Q1. 
   P'-P1' on $\Gamma''$ shows the same time interval as P-P1, but approaching $O$ with blueshifted photons (note: $\theta$ is now -ve). }
   \label{fig:doppler}
\end{figure}

The photon leaving $P$ may be considered to coincide with the wave crest of an EM emission of frequency $\nu'$ in the $\Gamma'$ frame. 
Let $P1$ be a second photon released at the next wave crest at a time $\delta{}t'$ later (Fig.\ref{fig:doppler}). 
This photon follows its own geodesic to arrive at $Q1$, at a time $\delta{}t_0$ later than $Q$ 
along the $\Gamma_0$ axis, with an observed frequency $\nu_0$. 
Thus the frequency of emission is $\nu'=1/{\delta{}t'}$ and the observed frequency of 
reception is $\nu_0=1/{\delta{}t_0}$.
%%--------------------
%%--------------------

Redshift is defined as:
\begin{equation}
z=(\nu'-\nu_0)/\nu_0=\nu'/\nu_0-1\,.
\label{eq:19}
\end{equation}
\begin{equation}
\textnormal{i.e. }z+1=\nu'/\nu_0=\delta{}t_0/\delta{}t'\,.
\label{eq:20}
\end{equation}
But for uniformly co-moving reference frames, $v$ and hence $\theta$ are constant. Hence, from~(\ref{eq:18}):
\begin{equation}
z+1=\delta{}t_0/\delta{}t'=t_0/t'=e^{\theta}\,.
\label{eq:21}
\end{equation}
And from Euler's formula, $e^{\theta}=\cosh{\theta}+\sinh{\theta}$; hence from (\ref{eq:12}) and (\ref{eq:13}):
\begin{equation}
z+1=\gamma+\gamma\beta=\gamma(1+\beta)\,.
\label{eq:22}
\end{equation}
This clearly approaches $\infty$ as $v\rightarrow{}c$. 
It may similarly be shown that reference frames approaching $\Gamma_0$ in Fig.~\ref{fig:doppler}, corresponding to an angle $-\theta$, 
are blue shifted by an amount $\gamma(1-\beta)$, which approaches 0 as $v\rightarrow{}c$, leading to (\ref{eq:23}):
\begin{equation}
z+1=\gamma(1\pm\beta)\,.
\label{eq:23}
\end{equation}
which is the relativistic Doppler shift. This may be compared with the classical Doppler shift of 
$(\nu-\nu_0)/\nu_0=1\pm\beta$, which it approaches in the limit of small $v$.

Using $\gamma=1/(1-\beta^2)^{1/2}$, an equivalent way of stating redshift is:
\begin{equation}
z=\gamma(1+\beta)-1=\left(\frac{1+\beta}{1-\beta}\right)^{1/2}\,.
\label{eq:24}
\end{equation}
or in terms of $v$:
\begin{equation}
v=c\left(\frac{(1+z)^2-1}{(1+z)^2+1}\right)\,.
\label{eq:25}
\end{equation}
The hyperbolic functions are not periodic, therefore the photon path of Fig.~\ref{fig:geodesic} can be continued to spiral indefinitely 
round the origin as $\theta$ is increased. For example, $\theta=360^{\circ}$ corresponds to $z=e^{2\pi}-1=534$, and 
$\beta=\tanh{2\pi}=0.999993$. The geometry remains correct and consistent, but the virtue of this model for visualization is lost at 
these high relativistic velocities.

%%%%%%%%%%%%%%%%%%%%%%%%%%%%%%%%%%%%%%%%%%%%%%%%%%%%%%%%
\subsection{Classical and relativistic Doppler redshift}
\label{section:doppler}
%%--------------------
\begin{figure}[ht]
   \centering
   \includegraphics[width=10cm]{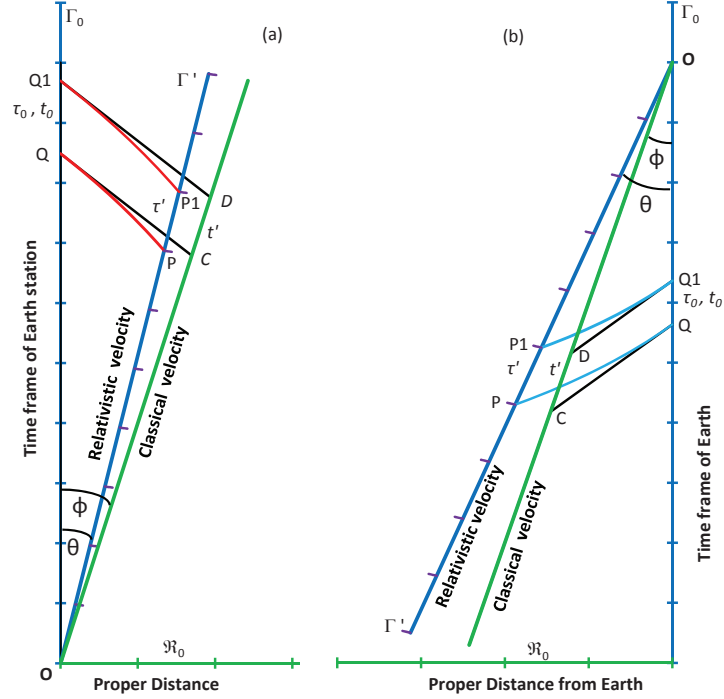}
   \caption{(a)~Relativistic Doppler redshift with successive photons emitted from $P$, $P1$ (blue line at $\theta$)
and observed at $Q$, $Q1$ and corresponding classical redshift 
with photons emitted from $C$ and $D$ (green line at $\phi$). 
(b)~Relativistic Doppler blueshift (blue line at $\theta$) with successive photons emitted 
and observed as in (a) and corresponding classical blueshift (green line at $\phi$).}
   \label{fig:redshift}
\end{figure}
%%--------------------
The difference between classical and relativistic velocities is demonstrated in Fig.~\ref{fig:redshift}, 
which superimposes the real $R-T_0$ axes over the imaginary $R-\Gamma_0$  ones. 
The real values $t'$ and $t_0$ for the classical velocity of the emitting object (green line) 
now correspond to the imaginary relativistic photon emission and reception intervals $\tau'$ and $\tau_0$. 
Let us assume a classical velocity of $0.25c$ (necessarily large to demonstrate the effect); 
then $\phi=\atan{0.25}\simeq14^{\circ}$. An observer on $\Gamma_0$ is assumed to know the (fixed) frequency of emission $\nu'={1}/{t'}$
and will measure the frequency of reception $\nu_0={1}/{t_0}$. 
From (\ref{eq:20}) and by the geometry of similar triangles, it may be shown that:
\begin{equation}
z=\frac{QQ1}{CD}-1=\frac{OQ}{OC}-1
\label{eq:26}
\end{equation}
\begin{equation}
\textnormal{and }\widehat{OCQ}=\widehat{ODQ1}=135^{\circ}-\phi
\label{eq:27}
\end{equation}
Hence using the sine rule, 
\begin{equation}
z=\cos\phi+\sin\phi-1\simeq 0.213
\label{eq:28}
\end{equation}
But because the transmitter is moving with relativistic speed, 
the true relativistic velocity (substituting for $z$ in Equ.~\ref{eq:25}) is $0.1905c$, 
which will subtend an angle at $O$ of $\theta=\atanh{(0.1905)}\simeq11^{\circ}$ (blue line of Fig.~\ref{fig:redshift}a). 
The frequencies of transmission and reception remain, of course, the same,
but it will be noted that the relativistic pulses occur earlier than the classical pulses, 
and hence will appear to be phase shifted.

If the velocities are uniform and $\ll{c}$, this relativistic correction is small and constant 
and would be of little account. But with a spacecraft subject to heliocentric gravitational deceleration, 
the discrepancy between the relativistic and classical velocities changes with distance and velocity, 
and will diminish with time, and hence appear as a small additional deceleration. 
The calculation of this discrepancy must include the effect of the varying flight trajectory 
projected towards the receiving station on Earth, and while it does not have an exact analytical solution, 
this is comparatively straightforward using computational iteration. 
Thus a spacecraft in hyperbolic trajectory at a geocentric distance of 21~A.U. 
and assessed classical velocity of 20~km/s will show a small additional component of acceleration 
of $\simeq8.4 \times 10^{-8}$~cm/s/s directed towards the observing station. 
This relativistic term appears as an additional geocentric deceleration 
that must be corrected for. It should be emphasised that, although this is of a similar magnitude 
to the unexpected solar-directed acceleration observed with the Pioneer craft,\cite{Anderson2002,Nieto2005} 
the SR correction distinctively decreases with distance and reducing velocity 
while the Pioneer anomaly was a constant heliocentric acceleration caused by 
anisotropic emission of thermal radiation from the Radioisotope Thermoelectric Generators (RTGs).\cite{Turyshev2012} 
%%%%%%%%%%%%%%%%%%%%%%%%%%%%%%%%%%%%%%%%%%%%%%%%%%%%%%%%%
\subsection{Relativistic and classical Doppler blueshift}
A similar effect is shown when an emitting object is approaching the Earth, 
but now the photons are blueshifted (Fig.~\ref{fig:redshift}b). Again, the blue line is 
the relativistic velocity and the green line the classical velocity, 
and again the imaginary and real axes are overlain for the periods of emission ($\tau'$ and $t'$) 
and reception ($\tau_0$ and $t_0$) and the corresponding frequencies remain identical, 
but now $z=\cos{\phi}-\sin{\phi}-1$ and hence the true (relativistic) velocity is here greater than the classical velocity. 
The true velocity of a spacecraft approaching 
a receiving station on Earth will therefore appear to accelerate compared to an assumed initial classical velocity and this may mimic the flyby effect, 
although again the effect will be very small, and may be highly variable as the relative velocities 
change with the rotation of the earth and its motion round the sun.\cite{Anderson2008}

%%%%%%%%%%%%%%%%%%%%%%%%%%%%%%%%%%%%%%%%%%%%%%%%%%%%%%%%%
\section{Geometry of time dilation and space contraction}
The value of visualizing time dilation and space contraction using the logarithmic-spiral geometry 
may be demonstrated by imagining an idealized relativistic particle entering the upper atmosphere 
perpendicular to a laboratory on Earth.
%%--------------------
\begin{figure}[ht]
   \centering
   \includegraphics[width=8cm]{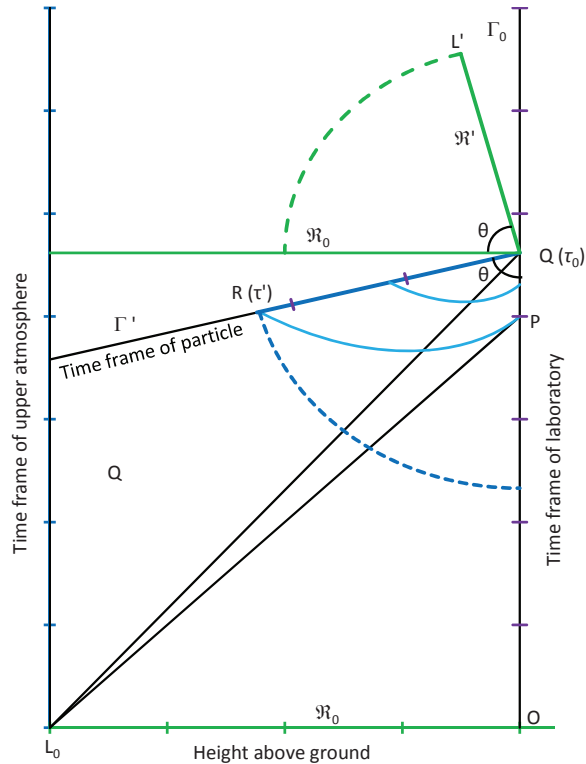}
   \caption{Time dilation and length contraction for a relativistic particle 
   descending vertically from the upper atmosphere (height $L_0$) to the Earth. $OQ$ is the world-line of the laboratory; $RQ$ 
   is the world-line of the particle. The blue curved lines correspond to two emitted blueshifted photons. The dashed lines show the rotations of the time (blue) and space (green) axes of the particle onto the corresponding rest frames of the laboratory.} 
    \label{fig:atmosphere}
\end{figure}
%%--------------------
 
In Fig.~\ref{fig:atmosphere}, $\Gamma_0$ represents the time line of the laboratory in its reference frame $\Sigma_0$, 
starting at physical location and time $O$, and the time line at $L_0$ represents the upper atmosphere, 
at height $L_0$ above the laboratory. Let us consider the particle entering the atmosphere
at $L_0$ with constant relativistic velocity $v$. In the time frames of the atmosphere and the laboratory, 
a photon emitted at $L_0$ is shown as the  line $L_0P$, at $45^\circ$ to each; this is a straight line 
because they are not moving relative to each other.

The path of the particle in its own reference frame is the line $RQ$ on $\Gamma'$, and it also emits a photon as it enters the 
atmosphere, shown by the blue spiral $RP$ (a second photon is also shown to emphasise the blueshifts). 
Setting $\gamma=\cosh{\theta}=2$ 
to emphasise the effect, then $\theta=75.4^{\circ}$  as shown in the figure, and $v=\tanh{\theta}=0.866c$, 
with the point of intersection at $Q$ defined by the logarithmic spiral $RP$. 

Proper time is the time measured for a system at rest, i.e. in its own reference frame.
The position of $Q$ is easy to determine, because the line $L_0Q$ represents 
the proper velocity of the particle as measured in the frame $\Sigma_0$, 
therefore the proper time is $t_0=OQ=L_0/v\simeq4.62$ units, and angle $\widehat{L_0QO}=\atan{0.866}\approx41^\circ$. 
In the frame of reference of the particle ($\Gamma'$), its own proper time is 
 $t'=RQ=PQ\times \exp{(-\theta)}=(t_0-OP)\times\exp{(-\theta)}\simeq2.31$ units, i.e.  
the time of flight as measured in the laboratory frame is twice as long as time measured 
in the frame of the particle, as expected with $\gamma=2$. 
The emission of the photons represents the same event at $L_0$ and $R$, but any idea of simultaneity is clearly lost.

The moving particle in its own frame of reference is, by definition, stationary. 
However, it observes the laboratory moving past it with a velocity $v$ over its own timeline of $t'$, 
and it asserts the laboratory moves a distance $L' = vt'=0.866\times 2.31=2$. 
Now the observer in the laboratory also agrees that the particle moves past her with velocity $v$, 
for this is an essential feature of relativity -- both observers agree about their relative velocity, 
only the sign changes. So she states the particle moves a distance $L_0=vt_0=0.866\times 4.62=4$. 
\begin{equation}
\textnormal{Hence: }L_0/L'=t_0/t'=2 \textnormal{ or } L'=L_0 / \gamma
\label{eq:29}
\end{equation}
%%--------------------
\begin{figure}[ht]
   \centering
   \includegraphics[width=8cm]{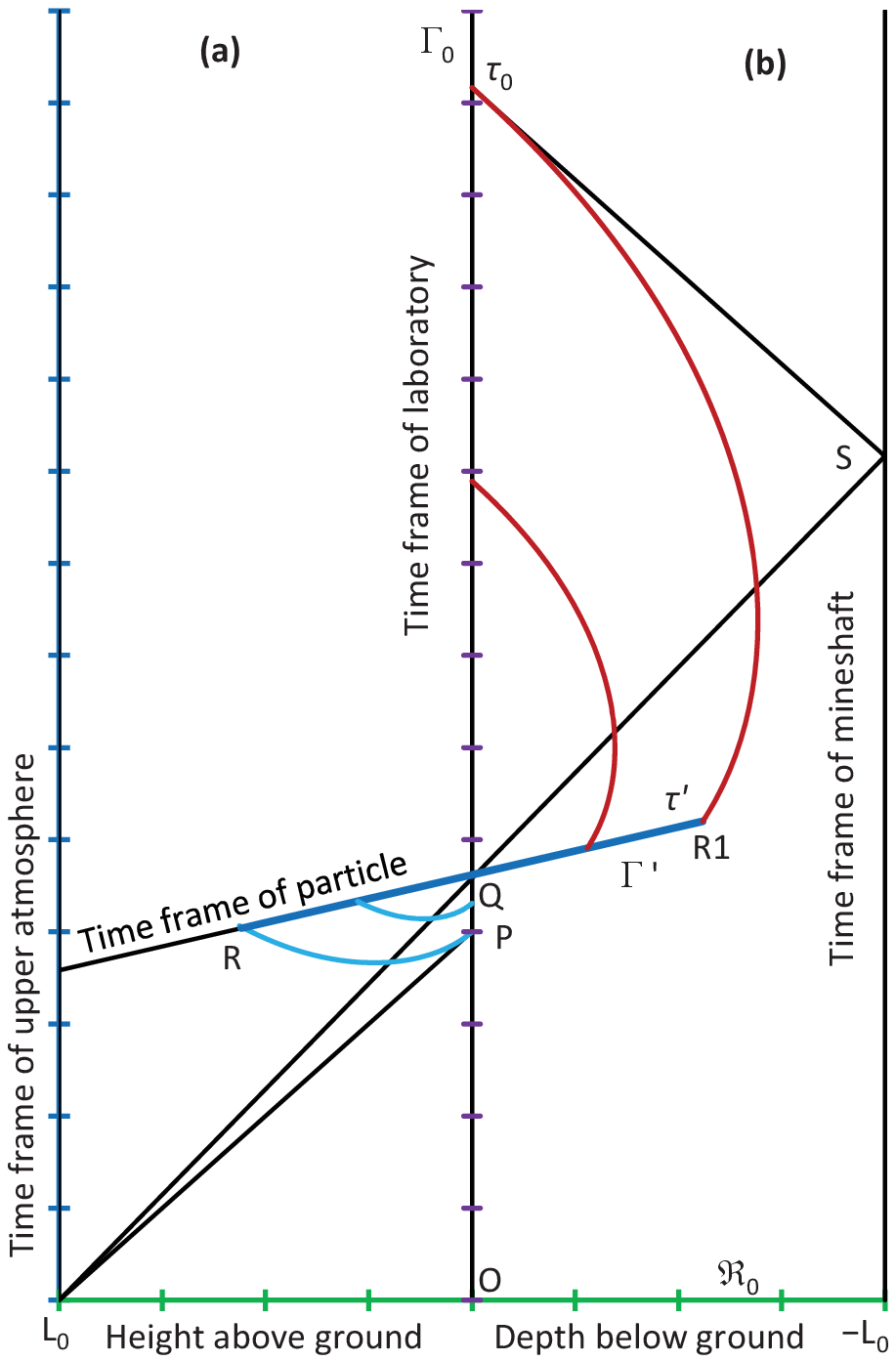}
   \caption{Time dilation and length contraction for a particle moving through the atmosphere and continuing down a deep mineshaft.}
    \label{fig:mine_shaft}
\end{figure}
%%--------------------
which is the Lorentz contraction relation of SR. These relationships are emphasized in Fig.~\ref{fig:atmosphere} 
by rotating the space and time coordinates of the particle frame back onto the corresponding space and time axes 
of the laboratory frame, shown as the dashed green and blue projection lines respectively. 

These relationships have been well verified in particle accelerator laboratories, 
where many studies of the decay times of both pions and mesons have compared their half-lives at rest 
in the laboratory to those measured when they attain relativistic velocities.\cite{Greenberg 1969,Bailey 1977} 
A precise experiment by Frisch and Smith measured the mean muon velocities 
and differences in count rates between their laboratory in Cambridge, Massachusetts, 
and Mount Washington at a height of 2000~m, allowing for slowing due to interaction with 
the atmosphere.\cite{Frisch 1963} This distance should be traversed by the muons in $6.4~\mu{}s$, 
but muons have a half-life of $2.2\mu sec$, and even at the speed of light 
this would give a range of only $660 m$. They measured a time dilation factor of $8.8\pm{}0.8$, corresponding to a mean velocity 
of $0.9934\pm{}0.0012c$ over this distance. This factor, with $\gamma\simeq30$, compresses $\tau'$ 
and the photon curves too much to demonstrate in Fig.~\ref{fig:atmosphere}, but the underlying geometry 
and mathematics remain identical.

One question often asked by students is how a moving clock slows equally whether moving towards or away from 
the observer, and this is illustrated in Fig.~\ref{fig:mine_shaft}. The left side of the diagram is the same 
as Fig.~\ref{fig:atmosphere}, with the particle approaching the laboratory at $Q$; but on the right side, 
imagine a deep mineshaft under the laboratory, of the same depth $L_0$ as the atmosphere, 
and the particle continues down this with its velocity unchanged. This is shown by the 
line $L_0QS$, which is the path of the particle in the reference frame of the laboratory. 
The particle track $R$-$R1$ is symmetrical about $Q$, therefore the distance travelled 
from the perspective of the particle ($Q$-$R1=L_2$) will again appear foreshortened by the factor 
$1/\gamma$. At the bottom of the mineshaft, the particle emits a photon at $R1$ which now follows 
the red logarithmic spiral, and a photon is also emitted from the mineshaft when the particle strikes $S$; both photons are detected 
back on the laboratory's world-line at $\Gamma_0$. 
Again, the points $S$ and $R1$ represent the same event, but simultaneity is lost as they occur on 
two different world-lines. A second photon is shown emitted half way through the particle's journey, 
and now the photons are heavily red-shifted and their arrival in the laboratory is ``smeared out'' in time. 
However, from symmetry, the world-line of the particle does not change, and the  
particle's time and distance axes remain the same; the depth of the mineshaft is contracted by the same factor, 
and the clock on the particle appears to be slowed by the same factor.
%%%%%%%%%%%%%%%%%%%%%%%%%%
\section{The Twin Paradox}
\begin{figure}[ht]
   \centering
   \includegraphics[width=7cm]{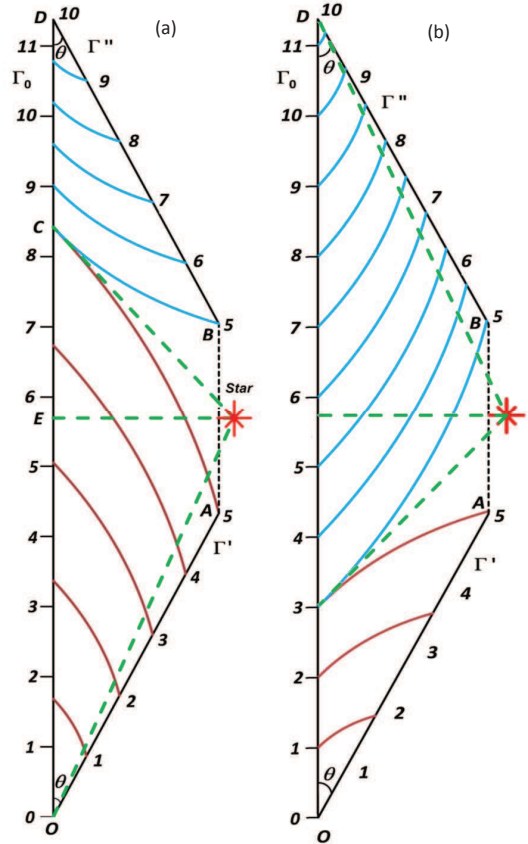}
   \caption{Geometry of the Twin Paradox: (a)~Twin~2 signalling to Twin~1; 
    (b)~Twin~1 signalling to Twin~2.}
    \label{fig:twins}
\end{figure}

In the twin paradox, it is customary to consider one twin remaining on the Earth while the second twin 
moves away for a certain time at a relativistic velocity $v$ before returning to Earth with the same velocity. 
To make a comparison between their relative times, we consider each twin to carry a standard clock 
which for convenience may be thought of as emitting a burst of photons every year, 
as measured in each clock's inertial frame. In Fig.~\ref{fig:twins}(a), Twin~1 stays on Earth and her clock ticks 
are shown on the vertical axis $\Gamma_0$. The point $S$, labelled $Star$, represents a distant object, 
assumed to be stationary w.r.t. Earth, hence its proper time runs parallel to that of Earth, 
and it is located at a proper distance $ES$, 2.7~light-years from Earth. 
Photons from $Star$ will therefore move along lines parallel to $SC$. 
Twin~2 leaves Earth at $O$ to follow the line $\Gamma'$, travelling for 5~years by her clock to 
point $A$. The angle $\theta$ has been chosen to be $30^{\circ}$, and the relativistic velocity for 
this angle may be obtained from $\beta=\tanh{30^{\circ}}$ or $v=0.48c$, shown by the line 
$OS$ in the Earth's proper time and length coordinates. The photon bursts from Twin~2 are shown crossing 
the dividing space at 1~year intervals to arrive on $\Gamma_0$ and be observed by Twin~1, 
where they may be compared with her clock. 

%%--------------------
%%--------------------
Twin~1 sees the red-shifted photons arriving at spaced-out intervals, as in Fig.~\ref{fig:twins}(a). 
However at time $C$, she notices that the photons are suddenly blue shifted and coming at more frequent intervals, 
until her twin returns to Earth and they compare clocks at position $D$. 
This is indicated in the figure by the photon line leaving $\Gamma''$ from $B$ at  $45^{\circ}$ and 
curving towards $\Gamma_0$, where they again arrive at $45^{\circ}$. 
It should be noted that this is a purely 
geometrical construct; the only requirement is that the geodesics follow lines of constant angle to the 
relevant inertial frames. But in order for successive photons not to cross paths, 
the geometry demands that the path of Twin~2 is broken, as shown by the broken line from $A$ to $B$. 
This illustrates the fact that Twin~2 must change direction (i.e. undergo acceleration) when her velocity reverses 
to a new world-line. The three points, $A$, $B$ and $Star$ all represent the same event in three inertial frames, 
again illustrating the lack of simultaneity. From the geometry of SR alone it may be seen that, 
while for Twin~2 a total of 10~years has passed, Twin~1 measures an elapsed time of 11.4~years, 
which may readily be confirmed from (\ref{eq:30}):
\begin{equation}
\gamma=\sqrt{1-v^2/c^2}=\cosh{30^{\circ}}=1.14
\label{eq:30}
\end{equation}
Therefore, as $t=t'\gamma$,  $t=11.4$~years.

The detailed geometry of the acceleration can only be considered under GR, but the net effect is 
a rapid rotation and shift of the inertial frame. In this model it represents an instantaneous change 
in velocity shown by the red- and blue-shifted photons arriving at the same time point, 
whereas a finite acceleration would smear these out over a finite time interval. It is, however, important 
to realise that the time dilation effect occurs equally on each leg and is independent of both 
the acceleration used to reverse the motion, and the Doppler shift of the photons. 
Both twins register 5~ticks of Twin~2's clock over the same length of journey, 
to $Star$ and back.

For comparison, the complementary view is shown in Fig.~\ref{fig:twins}(b), where Twin~1 sends yearly pulses 
to Twin~2. Twin~2 will now receive only 3 red-shifted pulses 
from Twin~1, but will receive over 8 blue-shifted pulses. Despite these differing views of each other's clocks, 
the total time dilation of Twin~1 to Twin~2 is still, of course, 11.4~years. 

%%%%%%%%%%%%%%%%%%%%%%%%%%%%%%%%%%%%%
\section{Relativistic radar distance}
%%--------------------
\begin{figure}[ht]
   \centering
   \includegraphics[width=5cm]{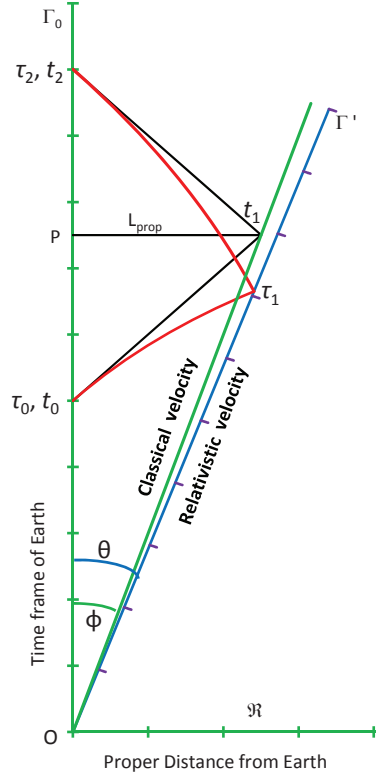}
   \caption{Relativistic radar distance (blue line at $\theta$) and corresponding classical (or proper) radar distance 
   (green line at $\phi$).}
    \label{fig:radar}
\end{figure}
%%--------------------
Looking at the spatial coordinate of $O$ (Fig.~\ref{fig:radar}), consider a rigid rod of proper length $L_{prop}$, 
moving through spacetime uniformly. A signal sent from $\Gamma_0$ at $t_0$ will propagate along the light cone 
to intercept $L_{prop}$'s world line at a time $t_1$. If the signal is reflected back towards 
the worldline of $\Gamma_0$ to arrive at time $t_2$, we may define $L_{prop}$ as the radar distance, such that
$L_{prop}=c(t_2-t_0)/2=c\Delta{t}$, and
\begin{equation}
v_{prop}=L_{prop}/OP=c\Delta{t}/(t_0+\Delta{t}) 
\label{eq:31}
\end{equation}
\begin{equation}
\textnormal{Hence }v_{prop}/c=\Delta{t}/(t_0+\Delta{t})=\tan\phi
\label{eq:32}
\end{equation}
By analogy with the proper radar distance, we may draw the geodesics for a photon leaving $\Gamma_0$ at time $\tau_0$, 
arriving at $\Gamma'$ at $\tau_1$ and being reflected back to return to $\Gamma_0$ at the received 
time $\tau_2$ (Fig.~\ref{fig:radar}). The radar distance is again defined as 
\begin{equation}
L_{rel}=c(t_2-t_0)/2 
\label{eq:33}
\end{equation}
and using (\ref{eq:18}) and (\ref{eq:21}): $\tau_1/\tau_0=e^{\theta}$; $\tau_2/\tau_1=e^{\theta}$
\begin{equation}
\textnormal{i.e. } \tau_2/\tau_0=t_2/t_0 = e^{2\theta}=(1+z)^2 
\label{eq:34}
\end{equation}
Substituting in (\ref{eq:33}) for $t_2$:
\begin{equation}
L_{rel}=ct_0[(1+z)^2-1]/2 
\label{eq:35}
\end{equation}
To bring real values back into the model, a value for $\theta$ has been chosen such that the 
time on the $\Gamma_0$ line is doubled for each reflection back. 
For this condition to be fulfilled, from (\ref{eq:34}):
\begin{equation}
t_2/t_0=(1+z)^2=2 
\label{eq:36}
\end{equation}
Hence $ z=\sqrt{2}-1\simeq0.414$ and for this value 
$\theta=\log{\sqrt{2}}$~rads $\simeq19.9^{\circ}$, as shown in Fig.~\ref{fig:radar}. 
This suggests that if an object moving away from us with a redshift of $z = 0.414$ could somehow reflect back 
photons it had received from us (such as the radio echo from a space probe), 
then setting the time of departure as the zero point (strictly speaking, our intercept with 
the timeline of the probe projected back at the moment it reflects the photons), 
those photons would be received exactly twice as long after the time they were transmitted. 

Again, as described in Section~\ref{section:doppler} for the relativistic Doppler redshift, 
for a system that is subject to deceleration relative to a receiving station on Earth, 
any phase-shift in the transmissions will occur earlier and the difference between proper and relativistic 
velocities will decrease with distance. In critical cases, or where measurements 
have a high sensitivity, this apparent deceleration may need to be accounted for.
%%%%%%%%%%%%%%%%%%%%%%%%%
\section{Discussion}
Following Einstein's original work in 1905,\cite{Einstein 1905} our understanding of SR goes back to Minkowski\cite{Minkowski 1909} and the four-vector formalism of Sommerfeld,\cite{Sommerfeld 1910a,Sommerfeld 1910b} and these form the basis for the geometric interpretation of SR in many contemporary presentations and textbooks. More recently, computerization has generated detailed images of how the world might appear to a relativistic traveller, and Weiskopf has provided a comprehensive review of the current status of visualization in SR, describing three visualization approaches: Minkowsi diagrams, spatial slices, and virtual camera models.\cite{Weiskopf 2010}   
Minkowski diagrams are the standard way of depicting the spacetime of SR graphically, by representing both temporal and spatial dimensions in a single image.  Spacetime events are visualized as points, the world-lines of objects as lines, and reference frames are indicated by their respective coordinate axes. The light rays emitted by an object and absorbed by the observer form a light cone that is also illustrated by a line, or by a collection of lines to generate a 3--D cone rendered as a 2--D image.

It is well recognised that problem-solving ability can be improved through the appropriate use of visualization,\cite{Anderson 1993,Waisel 1998} and the popularity of Minkowski diagrams in textbooks and other scientific representations of SR is rooted in the fact that those diagrams provide a geometric visualization of spacetime. Yet although SR has been fully vindicated since its first description by Einstein, and the underlying mathematical treatment is complete and self-consistent, the standard Minkowski representations have remained unchanged for over 100~years and their concepts remain difficult for some students to visualize.\cite{Tanel 2014} 

This paper presents a novel geometrical picture of SR in which the constancy of $c$ to any observer is taken as axiomatic. This is represented by a geodesic describing a constant angle to a set of diverging reference frames; the locus of such a line is the logarithmic spiral on the imaginary plane, which is a self-similar curve whose relationships are independent of scale or the angle of rotation. It must be emphasised that the visualization of the photon path in this model is a mathematical abstraction, and must not be construed as representing the actual location of the photons. In practice, we can only know the velocity of the emitting frame ($\Gamma'$) at the moment of emission relative to our own frame ($\Gamma_0$) at the moment of reception. The actual position of the photons between these two moments is indeterminate (as demanded by quantum theory), and the path as a logarithmic spiral can only be defined and constructed once the relative velocities at these times is defined and stated.
This geometry, however, is self-consistent, leading naturally to the common mathematical relationships of SR, and this is developed to provide an alternative geometrical representation to the standard Minkowski diagrams. Relativistic Doppler redshift, radar distance, and the twin paradox are presented as examples of this geometrical approach.

The light cone is shown to have relevance only in the local or ``proper'' frame of reference with ``real'' time, where velocities are given in terms of $\tan\phi$ and therefore cannot exceed $45^{\circ}$ at the surface of the cone, whereas the velocities of independent inertial frames are defined on the imaginary plain in terms of $\tanh\theta$ and may take any angle. This difference between proper and relativistic velocities is emphasised, and it is shown that, where these velocities are changing, this may be revealed as a small apparent acceleration or deceleration, even in systems not generally thought to be moving at relativistic velocities. 

One problem with the logarithmic spiral visualization is the reduction of dimensionality, as only the temporal dimension and one spatial dimension are shown in the diagrams. However, all approaches have in common a reduction of dimensionality of 4-D spacetime, and Minkowski diagrams also never show all four dimensions of spacetime but only a restricted subspace. Another issue is the interpretation of angles since the Minkowski metric in imaginary space is different from the Euclidean metric, and the hyperbolic functions cannot be inferred intuitively from our experience with Euclidean space in which the diagram is rendered. For example, the maximum angle available in the local frame of reference is $\atan{}(1)=45^\circ$ for the photons; but for the diverging reference frames, the maximum angle is $\atanh{}(1)=\infty$; hence the geodesic can spiral round the origin indefinitely as the relative velocity of the frames draws closer to $c$. Also, the length of the photon path is the null geodesic which appears to have finite length on the diagrams, but as Weiskopf has stated, this interpretation issue is equally a problem with the standard Minkowski diagrams, which assume a mathematical background sufficient to make them intelligible.\cite{Weiskopf 2010} 

Einstein is quoted as saying, ``My ability lies not in mathematical calculation, but rather in visualizing effects, possibilities and consequences'',\cite{Pinker 1990} and visualization using computer-generated graphics can be a means of finding simplicity in a complex artefact by the selective representation of an appropriate abstraction.\cite{Petre 1998,Zagami 2012} The logarithmic spiral geometry of SR is an alternative visual representation for understanding SR as pure geometry that might complement the standard Minkowski diagrams. Using computer-generated diagrams to show the photon as an equiangular locus, or logarithmic spiral, is supplementary to Minkowski diagrams, and demonstrates how $c$ can remain constant to each observer, while simultaneously displaying the classical and relativistic time frames. This may enable the student to understand the differences between these reference frames, and the consequences of varying the relativistic velocity in a variety of scenarios.

%%%%%%%%%%%%%%%%%%%%%%%%%%
\section*{Acknowledgments}
I would like to thank the anonymous referee for many helpful suggestions and comments.
%%%%%%%%%%%%%%%%%%%%%%%%%%
%%\section*{References}

\end{document}